\documentclass[11pt]{article}
\usepackage{amssymb, amsmath}
\usepackage{graphics}
\usepackage[dvips]{graphicx}
\usepackage{eepic}
\usepackage{epic}

\usepackage{array}
\usepackage{amscd}
\usepackage{subfigure}
\newtheorem{thm}{Theorem}[section]

\newtheorem{prp}[thm]{Proposition}

\newcommand{\DIS}{\displaystyle}

\def\R{{\mathbb R}}

\def\nn{{\nonumber}}

\def\bI{\text{\mathversion{bold}{$I$}}}
\def\bV{\text{\mathversion{bold}{$V$}}}
\def\bbI{\text{\mathversion{bold}{$\bar I$}}}
\def\bbV{\text{\mathversion{bold}{$\bar V$}}}
\def\bJ{\text{\mathversion{bold}{$J$}}}
\def\bW{\text{\mathversion{bold}{$W$}}}
\def\bbJ{\text{\mathversion{bold}{$\bar J$}}}
\def\bbW{\text{\mathversion{bold}{$\overline W$}}}

\title{\textbf{{A geometric realization of the ultradiscrete periodic Toda lattice via tropical plane curves}}}
\author{Atsushi \textsc{Nobe}\footnote{Department of Mathematics, Faculty of Education, Chiba University, 1-33 Yayoi-cho Inage-ku, Chiba 263-8522, Japan.
\newline e-mail: \texttt{nobe@faculty.chiba-u.jp}
\newline This work was partially supported by JSPS KAKENHI Grant Number 22740100.}}
\date{}
\begin{document}
%

\maketitle

\begin{abstract}      
This is a review article on a tropical geometric realization of the ultradiscrete periodic Toda lattice (UD-pTL).
Time evolution of the UD-pTL is translated into an addition on the Picard group of its spectral curve, which is a tropical hyperelliptic curve of arbitrary genus depending on the system size.
The addition on the Picard group can be realized by using intersection of several tropical plane curves, one of which is the spectral curve. 
In addition, the tropical eigenvector map, which maps a point in the phase space of the UD-pTL into a set of points on the spectral curve, can also be realized by using intersection of tropical curves.
Thus, if the initial values are given then the time evolution of the UD-pTL is completely translated into a motion of intersection points of tropical plane curves.
Moreover, all tropical plane curves appearing in the curve intersection are explicitly given in terms of the conserved quantities of the UD-pTL.
\end{abstract}

\section{Introduction}
The discovery of a cellular automaton possessing solitonical nature by Takahashi -- Satsuma in 1990 \cite{TS90} was a trigger of various subsequent  studies upon new integrable dynamical systems governed by piecewise linear maps on the real spaces of finite or infinite dimension.
Such dynamical systems are called ultradiscrete integrable systems and their evolutions are described by means of the max-plus algebra.
The most powerful tool to investigate ultradiscrete integrable systems known as the procedure of ultradiscretization was proposed by Tokihiro et al. in 1996 \cite{TTMS96}.
By using this procedure, ultradiscrete integrable systems are directly connected with the discrete integrable systems such as the discrete KdV,  the discrete Toda and the discrete KP equations.
Namely, evolution equations, exact solutions and conserved quantities of such discrete integrable systems are simultaneously reduced to those of ultradiscrete integrable systems by applying ultradiscretization \cite{NT98,TTM00,IMNS04}.
With ultradiscretization as a booster, ultradiscrete integrable systems have intensively been studied by using various mathematical tool such as combinatorics \cite{MSTTT97,FOY00,YYT03,NY04}, crystal bases of quantum groups \cite{HIK99,HHIKTT01,HKOTY02,IKT12}, tropical geometry \cite{Nobe08, KNT08,KKNT09,Nobe11,IKT12} and so forth.

In this article, we focus our interests on the geometric aspects of ultradiscrete integrable systems.
The first step of the studies on the geometry of ultradiscrete integrable systems was made by  Kimijima -- Tokihiro in 2002 \cite{KT02}.
They applied the procedure of ultradiscretization to the periodic discrete Toda lattice (pdTL) and their quasi-periodic solutions, and linearized time evolution of the ultradiscrete periodic Toda lattice (UD-pTL) on the real torus corresponding to the Jacobian of the spectral curve of the pdTL.
They also solved the initial value problem to the UD-pTL of genus 1 and constructed its quasi-periodic solutions by means of the ultradiscrete elliptic theta functions.
Subsequent steps were made by Inoue -- Takenawa \cite{IT08,IT09} and Iwao et al. \cite{IIMT09,Iwao10} in terms of tropical geometry \cite{IMS07} and ultradiscretization procedure.
They solved the initial value problems to the UD-pTL of arbitrary genus by using tropical hyperelliptic curves, their tropical Jacobians, ultradiscrete theta functions and ultradiscretization of Abelian integrals.
Through their works, we have gradually recognized the advantage in applying the method of tropical geometry to ultradiscrete integrable systems. 

We will go a step further and try to realize the UD-pTL completely in the framework of tropical geometry.
In section \ref{sec:ATHCPBS} and \ref{sec:Addition}, we introduce the UD-pTL and its spectral curve and realize time evolution of the UD-pTL as an addition of points on the Picard group of its spectral curve.
We give explicit formulae for the eigenvectors of the Lax matrix of the pdTL in terms of the conserved quantities in section \ref{sec:TropicalEigenvectorMap}.
Moreover, we realize the tropical eigenvector maps geometrically by using intersection of the spectral curve of the UD-pTL and tropical plane curves derived from the eigenvector of the Lax matrix.
In section \ref{sec:AGROP}, we translate the addition of points on the Picard group into intersection of tropical plane curves.
These tropical plane curves, one of which is the spectral curve of the UD-pTL, are explicitly given by using the conserved quantities of the UD-pTL.
If we restrict the UD-pTL to take its initial values in positive integers then its time evolution is realized as a motion of balls in an array of boxes and is called the periodic box-ball system (pBBS) \cite{YT02}.
In \ref{sec:example}, we present a concrete computation of the geometric realization of the pBBS for a certain initial condition by using intersection of tropical plane curves.
We observe that the time evolution of the pBBS is realized as a motion of several points on a tropical hyperelliptic curve.

\section{Ultradiscrete periodic Toda lattice and periodic box-ball system}
\label{sec:ATHCPBS}
Let us consider the dynamical system $\zeta: \R^{2g+2}\to\R^{2g+2};$
\begin{align*}
(\bJ,\bW)
=
(J_1,\ldots,J_{g+1},W_1,\ldots,W_{g+1})
\mapsto
(\bbJ,\bbW)
=
(\bar J_1,\ldots,\bar J_{g+1},\overline W_1,\ldots,\overline W_{g+1})
\end{align*}
given by
\begin{align}
\bar J_i&=\left\lfloor W_i,X_i+J_i\right\rfloor,\qquad
\overline{W}_i=J_{i+1}+W_i-\bar J_i,\label{eq:pbbs}\\
X_i&=\left\lceil\sum_{l=1}^k\left(J_{i-l}-W_{i-l}\right)\right\rceil_{0\leq k\leq g}.\nn
\end{align}
Here we define
\begin{align*}
\lfloor A,B,\ldots\rfloor:=\min\left(A,B,\ldots\right),
\qquad
\lceil A,B,\ldots\rceil:=\max\left(A,B,\ldots\right)
\end{align*}
for $A,B,\ldots\in\R$.
Moreover, we assume that $\sum_{l=1}^0\left(J_{i-l}-W_{i-l}\right)=0$ and 
\begin{align}
\sum_{i=1}^{g+1}J_i<\sum_{i=1}^{g+1}W_i.
\label{eq:jwcond}
\end{align}
When we iterate the map $\zeta$ we use the notation
\begin{align*}
(\bJ^t,\bW^t)
=
\left(J_1^t,\ldots,J_{g+1}^t,W_1^t,\ldots,W_{g+1}^t\right)
:=
\underset{t}{\underbrace{\zeta\circ\zeta\circ\cdots\circ\zeta}}(\bJ,\bW)
\end{align*}
for $t=0,1,\ldots$.

We call the dynamical system generated by (\ref{eq:pbbs}) the ultradiscrete periodic Toda lattice (UD-pTL).
It is well known that the UD-pTL is reduced from the periodic discrete Toda lattice or the periodic discrete KdV equation through the procedure of {ultradiscretization} \cite{TTMS96}.
In particular, if the variables $J_1,\ldots,J_{g+1}$ and $W_1,\ldots,W_{g+1}$ take their values in positive integers then the dynamical system is called the periodic box-ball system (pBBS) \cite{YT02, KT02}.
The pBBS is obtained from the box-ball system (BBS), which was introduced by Takahashi -- Satsuma as a soliton cellular automaton \cite{TS90}, by imposing a periodic boundary condition.

\begin{figure}[htbp]
\centering
{\unitlength=.03in{\def\arraystretch{1.0}
\begin{picture}(150,35)(0,-5)
\thicklines
\multiput(0,0)(10,0){9}{\line(0,1){10}}
\put(0,0){\line(1,0){80}}
\put(0,10){\line(1,0){80}}


\multiput(100,0)(10,0){6}{\line(0,1){10}}
\put(100,0){\line(1,0){50}}
\put(100,10){\line(1,0){50}}

\dottedline(80,0)(100,0)
\dottedline(80,10)(100,10)

\multiput(5,5)(10,0){3}{\circle{8}}
\multiput(55,5)(10,0){1}{\circle{8}}
\multiput(115,5)(10,0){2}{\circle{8}}

\dashline{2}(0,-2)(30,-2)
\dashline{2}(50,-2)(60,-2)
\dashline{2}(110,-2)(130,-2)

\dashline{2}(30,12)(50,12)
\dashline{2}(130,12)(150,12)
\dashline{2}(60,12)(82,12)
\dashline{2}(98,12)(110,12)

\qbezier[60](150,5)(170,25)(75,25)
\qbezier[60](0,5)(-20,25)(75,25)
\put(15,-6){\makebox(0,0){$J_1$}}
\put(55,-6){\makebox(0,0){$J_2$}}
\put(120,-6){\makebox(0,0){$J_{g+1}$}}

\put(40,16){\makebox(0,0){$W_1$}}
\put(70,16){\makebox(0,0){$W_2$}}
\put(105,16){\makebox(0,0){$W_{g}$}}
\put(140,16){\makebox(0,0){$W_{g+1}$}}
\end{picture}
}}
\caption{Correspondence of $\bJ$ and $\bW$ with the variables of the pBBS.
}
\label{fig:pBBS}
\end{figure}
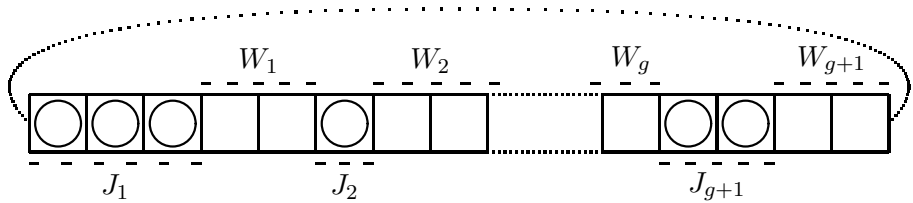

The pBBS can be realized by using an array of finite boxes and balls as follows.
Consider an array of $\sum_{i=1}^{g+1}\left(J_i+W_i\right)$ boxes whose both ends are connected each other.
Assume that we have $\sum_{i=1}^{g+1}J_i$ balls.
Put a ball into an arbitrary box in the array of boxes.
Successively, put a ball into the next box of the occupied one to the right.
Repeat this procedure $J_1$ times then we obtain a sub-array of consecutive $J_1$ boxes each of which is occupied with a ball.
For the adjoining $W_1$ boxes of the right-most occupied one to the right, we do not put balls into them.
Thus we obtain a sub-array of consecutive $W_1$ empty boxes adjoining the sub-array of $J_1$ occupied ones.
Next, put a ball into the next box of the right-most box in the sub-array of $W_1$ empty boxes to the right, and repeat the above procedure for $J_2$ and $W_2$.
Then we obtain a sub-array of consecutive $J_2$ occupied boxes and that of consecutive $W_2$ empty boxes adjoining to the sub-array of $W_1$ empty boxes.
By applying the procedure repeatedly, we finally obtain sub-arrays of $J_1, \ldots,J_{g+1}$ occupied boxes and those of $W_1,\ldots,W_{g+1}$ empty boxes which are alternately arranged (see figure \ref{fig:pBBS}). 
This is the initial state of the pBBS.

Time evolution \eqref{eq:pbbs} can be realized by moving balls in the following manner.
Make a copy of every ball in the initial state constructed above.
Pick an arbitrary copy of ball and move it to the nearest empty box to the right. 
Repeat this procedure till every ball moves once, and then erase the original ones.
Thus we obtain the second state consisting of $\sum_{i=1}^{g+1}W_i$ empty boxes and $\sum_{i=1}^{g+1}J_i$ occupied boxes each of which contains a ball.
We find that the sub-array of occupied boxes starting from the left-most box in that of $W_1$ empty boxes in the initial state has length $\bar J_1$.
We also find that the sub-array of empty boxes adjoining that of $\bar J_1$ occupied boxes has length $\overline W_1$.
We inductively find that the second state consists of sub-arrays of  $\bar J_1,\ldots,\bar J_{g+1}$ occupied boxes and those of $\overline W_1,\ldots,\overline W_{g+1}$ empty boxes which are alternately arranged (see figure \ref{fig:pBBSte}).
Repeat this procedure $t$ times then we obtain the $t$-th state of the balls corresponding to $\{\bJ^t, \bW^t\}$.

\begin{figure}[htbp]
\centering
{\unitlength=.03in{\def\arraystretch{1.0}
\begin{picture}(150,70)(0,-7)
{\thicklines{
\multiput(0,30)(10,0){9}{\line(0,1){10}}
\put(0,30){\line(1,0){80}}
\put(0,40){\line(1,0){80}}

\multiput(100,30)(10,0){6}{\line(0,1){10}}
\put(100,30){\line(1,0){50}}
\put(100,40){\line(1,0){50}}

\dottedline(80,30)(100,30)
\dottedline(80,40)(100,40)

\multiput(5,35)(10,0){3}{\circle{8}}
\multiput(55,35)(10,0){1}{\circle{8}}
\multiput(115,35)(10,0){2}{\circle{8}}
}}

\multiput(5,45)(10,0){3}{\circle{8}}
\multiput(55,45)(10,0){1}{\circle{8}}
\multiput(115,45)(10,0){2}{\circle{8}}


{\thicklines{

\multiput(0,0)(10,0){9}{\line(0,1){10}}
\put(0,0){\line(1,0){80}}
\put(0,10){\line(1,0){80}}

\multiput(100,0)(10,0){6}{\line(0,1){10}}
\put(100,0){\line(1,0){50}}
\put(100,10){\line(1,0){50}}

\dottedline(80,0)(100,0)
\dottedline(80,10)(100,10)

\multiput(35,5)(10,0){2}{\circle{8}}
\multiput(65,5)(10,0){2}{\circle{8}}
\multiput(135,5)(10,0){2}{\circle{8}}

\dashline{2}(30,-2)(50,-2)
\dashline{2}(60,-2)(80,-2)
\dashline{2}(130,-2)(150,-2)

\dashline{2}(0,12)(30,12)
\dashline{2}(50,12)(60,12)
\dashline{2}(98,12)(130,12)

\qbezier[80](15,50)(35,60)(35,14)
\put(35,14){\vector(0,-1){3}}
\qbezier[80](25,50)(45,60)(45,14)
\put(45,14){\vector(0,-1){3}}
\qbezier[80](5,50)(65,70)(65,14)
\put(65,14){\vector(0,-1){3}}
\qbezier[80](55,50)(75,60)(75,14)
\put(75,14){\vector(0,-1){3}}
\qbezier[80](125,50)(135,60)(135,14)
\put(135,14){\vector(0,-1){3}}
\qbezier[80](115,50)(145,70)(145,14)
\put(145,14){\vector(0,-1){3}}
\put(40,-6){\makebox(0,0){$\bar J_1$}}
\put(70,-6){\makebox(0,0){$\bar J_2$}}
\put(140,-6){\makebox(0,0){$\bar J_{g+1}$}}

\put(55,16){\makebox(0,0){$\overline W_1$}}
\put(115,16){\makebox(0,0){$\overline W_{g}$}}
\put(15,16){\makebox(0,0){$\overline W_{g+1}$}}
}}
\end{picture}
}}
\caption{Time evolution of the pBBS.
}
\label{fig:pBBSte}
\end{figure}

Now we introduce the spectral curve of the UD-pTL. 
Let us consider the following tropical polynomial $F$ in $X$ and $Y$
\begin{align*}
F(X,Y)
:=
\left\lfloor
2Y,Y+\left\lfloor(g+1)X,C_g+gX,\cdots,C_1+X,C_0\right\rfloor,C_{-1}
\right\rfloor,
\end{align*}
where $C_{-1},C_0,\ldots,C_g\in\mathbb{T}:=\R\cup\{\infty\}$. 
We can associate the coefficients $C_{-1},C_0,\ldots,C_g$ in $F$ with the variables $\bJ$ and $\bW$ of the UD-pTL appropriately \cite{MIT05,IT08,Nobe13}, e.g.,
\begin{align*}
&C_g
=
C_g(\bJ;\bW)
=
{\left\lfloor{J_i,W_i}\right\rfloor_{1\leq i\leq g+1}},\\
&C_{g-1}
=
C_{g-1}(\bJ;\bW)
=
\left\lfloor
\underset{1\leq i< j\leq g+1}{\left\lfloor{J_i+J_j,W_i+W_j}\right\rfloor},
\underset{\substack{1\leq i, j\leq g+1\\j\neq i,i-1}}{\left\lfloor J_i+W_j\right\rfloor}
\right\rfloor,\\
&C_0
=
C_0(\bJ;\bW)
=
\left\lfloor
\sum_{i=1}^{g+1}J_i,\sum_{i=1}^{g+1}W_i
\right\rfloor
=
\sum_{i=1}^{g+1}J_i,\\
&C_{-1}
=
C_{-1}(\bJ;\bW)
=
\sum_{i=1}^{g+1}\left(J_i+W_i\right).
\end{align*}
Here we use the assumption (\ref{eq:jwcond}).
The correspondence defines a piecewise linear map
\begin{align}
\psi:\R^{2g+2}\to\R^{g+2};\ 
\left(\bJ,\bW\right)
\mapsto
\left(C_{-1},C_0,\ldots,C_g\right).
\label{eq:CJW}
\end{align}
We may use the notation $C_i^t:=C_i\left(\bJ^t;\bW^t\right)$  for $t=0,1,\ldots$ when the map $\zeta$ is iterated.
One can show that the coefficients $C_{-1},C_0,\ldots,C_g$ in $F$ are the conserved quantities of the UD-pTL \cite{MIT05,IT08}.

\begin{figure}[t]
\centering
{\unitlength=.05in{\def\arraystretch{1.0}
\begin{picture}(80,64)(0,16)
\thicklines
\thicklines
\put(51,45){\line(-4,-1){8}}
\put(51,51){\line(-4,1){8}}
\put(51,45){\line(0,1){6}}
\put(43,43){\line(-3,-1){6}}
\put(43,53){\line(-3,1){6}}
\put(43,43){\line(0,1){10}}
\put(37,41){\line(0,1){14}}

\dashline{1}(37,41)(27,36)
\dashline{1}(37,55)(27,60)

\put(27,36){\line(0,1){24}}
\put(27,36){\line(-3,-2){6}}
\put(27,60){\line(-3,2){6}}
\put(21,32){\line(0,1){32}}
\put(21,32){\line(-1,-1){8}}
\put(21,64){\line(-1,1){8}}
\put(13,24){\line(0,1){48}}

\dottedline(66,45)(51,45)
\dottedline(66,51)(51,51)

\dottedline(13,24)(9,16)
\dottedline(13,72)(9,80)
\put(15,22){\makebox(0,0){$V_0^\prime$}}
\put(22,29){\makebox(0,0){$V_1^\prime$}}
\put(51,42){\makebox(0,0){$V_g^\prime$}}
\put(15,73){\makebox(0,0){$V_0$}}
\put(22,66){\makebox(0,0){$V_1$}}
\put(51,54){\makebox(0,0){$V_g$}}
\end{picture}
}}
\caption{
The tropical hyperelliptic curve $\tilde\Gamma$.
The compact subset $\Gamma$ is drown by solid lines and the half rays by dotted lines.
}
\label{fig:THC}
\end{figure}
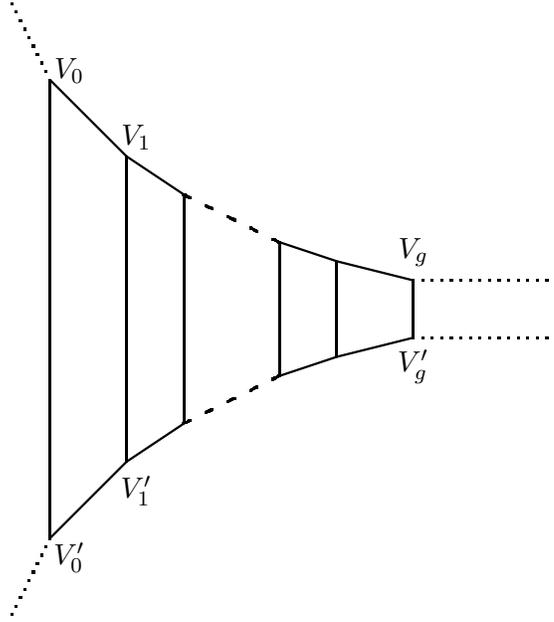

Consider the tropical plane curve $\tilde\Gamma$ defined by $F$ \cite{RGST03,Gathmann06,IMS07}:
\begin{align*}
&\tilde\Gamma
:=
\left\{
P\in\R^2\ |\ 
\mbox{$F$ is not differentiable at $P$}
\right\}.
\end{align*}
Assume $C_{-1}>2C_0$, $C_{g-1}>2C_g$, and $C_i+C_{i+2}>2C_{i+1}$ for $i=0,1,\ldots,g-2$.
Then $\tilde\Gamma$ is a tropical hyperelliptic curve of genus $g$ \cite{MZ06,HMY09}.
By removing all half rays from $\tilde\Gamma$, we obtain the compact tropical curve denoted by $\Gamma$ (see figure \ref{fig:THC}).
Hereafter, we consider $\Gamma$ rather than $\tilde\Gamma$ as the spectral curve of the UD-pTL.

Consider the involution $\iota:\Gamma\to\Gamma; P=(X,Y)\mapsto P^\prime=(X,C_{-1}-Y)$.
We call $P^\prime$ the conjugate of $P$.
Note that $\Gamma$ is symmetric with respect to the line $Y={C_{-1}}/{2}$.
We denote the $2g+2$ vertices of $\Gamma$ by $V_i$ and $V_i^\prime$ for $i=0,1,\ldots,g$
\begin{align*}
V_i
&=
\left(
C_{g-i}-C_{g-i+1}, C_{-1}-(g-i+1)C_{g-i}+(g-i)C_{g-i+1}
\right),\\
V_i^\prime
&=
\left(
C_{g-i}-C_{g-i+1}, (g-i+1)C_{g-i}-(g-i)C_{g-i+1}
\right).
\end{align*}
The cycle connecting $V_i$, $V_{i-1}$, $V_{i-1}^\prime$, and $V_i^\prime$ in a counterclockwise direction is denoted by $\alpha_i$ for $i=1,2,\ldots,g$.

\section{Addition on tropical hyperelliptic curves}\label{sec:Addition}
We briefly review addition of points on tropical hyperelliptic curves \cite{Vigeland04,Nobe12}.

Denote the divisor group of $\Gamma$ by $\mathcal{D}(\Gamma)$.
A rational function on $\Gamma$ is a continuous function $f:\Gamma\to\R$ such that its restriction to any edge is piecewise linear with integral slope \cite{Gathmann06}.
The order of $f$ at $P\in\Gamma$ is the sum of the outgoing slopes of all segments emanating from $P$ and is denoted by ${\rm ord}_Pf$.
If ${\rm ord}_Pf<0$ then $P$ is called the zero of $f$ of order $\left|{\rm ord}_Pf\right|$.
If ${\rm ord}_Pf>0$ then $P$ is called the pole of $f$ of order ${\rm ord}_Pf$.
The principal divisor $(f)$ of $f$ is defined to be $(f):=\sum_{P\in\Gamma}\left({\rm ord}_Pf\right)P$.
We then find $\deg (f)=\sum_{P\in\Gamma}\left({\rm ord}_Pf\right)=0$.

Define the Picard group of $\Gamma$ to be the residue class group ${\rm Pic}^0(\Gamma):=\mathcal{D}_0(\Gamma)\slash\mathcal{D}_l(\Gamma)$, where $\mathcal{D}_0(\Gamma)$ is the group of divisors of degree 0 on $\Gamma$ and $\mathcal{D}_l(\Gamma)$ is the group of principal divisors of rational functions on $\Gamma$.

Define the canonical map $\Phi:\mathcal{D}_g^+(\Gamma)\to {\rm Pic}^0(\Gamma)$ to be
\begin{align*}
\Phi(A)
:\equiv
A-D^\ast
\quad
\mbox{(mod $\mathcal{D}_l(\Gamma)$)}
\qquad
\mbox{for $A\in\mathcal{D}_g^+(\Gamma)$},
\end{align*}
where $\mathcal{D}_g^+(\Gamma)$ is the group of effective divisors of degree $g$ on $\Gamma$ and $D^\ast\in\mathcal{D}_g^+(\Gamma)$ is a fixed element.
We then have the following theorem.
\begin{thm}[\cite{Nobe12}]
The canonical map $\Phi$ is surjective.
In particular, $\Phi$ is bijective if $g=1$.
\end{thm}

By using the surjection $\Phi$, we induce an addition of points on the $g$-th symmetric product ${\rm Sym}^g(\Gamma):=\Gamma^g/\mathfrak{S}_g$ from ${\rm Pic}^0(\Gamma)$.
Put $\tilde\Phi:=\Phi\circ\mu^{-1}:{\rm Sym}^g(\Gamma)\to{\rm Pic}^0(\Gamma)$, where $\mu:\mathcal{D}_g^+(\Gamma)\to{\rm Sym}^g(\Gamma); D_P=P_1+P_2+\cdots+P_g\mapsto d_P:=\mu(D_P)=\left\{P_1,P_2,\ldots,P_g\right\}$.
For $d_P, d_Q\in{\rm Sym}^g(\Gamma)$, we define $d_P\oplus d_Q$ to be an element in the subset
\begin{align*}
\tilde\Phi^{-1}\left(\tilde\Phi(d_P)+\tilde\Phi(d_Q)\right)
\subset{\rm Sym}^g(\Gamma).
\end{align*}
Although $d_P\oplus d_Q$ is not uniquely determined on ${\rm Sym}^g(\Gamma)$, there exists a subset of ${\rm Sym}^g(\Gamma)$ on which the addition is uniquely determined.

Put $\alpha_{ij}:=\alpha_i\cap\alpha_j\setminus\left\{\mbox{end points of $\alpha_i\cap\alpha_j$}\right\}$ for the cycles $\alpha_i$ ($i=1,2,\ldots,g$).
We define the subset $\tilde{\mathcal{D}}$ of $\mathcal{D}_{g}^{+}(\Gamma)$ to be
\begin{align*}
\tilde{\mathcal{D}}
:=
\left\{
\begin{array}{l}
D_P\in\mathcal{D}_{g}^{+}(\Gamma)\\
\end{array}
\
\left|
\begin{array}{l}
\mbox{$P_i\in\alpha_i$ for all $i=1,2,\ldots,g$ and} \\
\mbox{there exists at most one point on $\alpha_{ij}$}
\end{array}
\right.
\right\}.
\end{align*}
We then have the following theorem.
\begin{thm}[\cite{IT08}]\label{thm:bijection}
The reduced map $\Phi|{\tilde{\mathcal{D}}}:\tilde{\mathcal{D}}\overset{\sim}{\to}{\rm Pic}^0(\Gamma)$ is bijective.
\end{thm}
Since $\mu|\tilde{\mathcal{D}}:\tilde{\mathcal{D}}\to\mu(\tilde{\mathcal{D}})\subset{\rm Sym}^g(\Gamma)$ is also bijective, the addition $\oplus:\mu(\tilde{\mathcal{D}})\times \mu(\tilde{\mathcal{D}})\to\mu(\tilde{\mathcal{D}})$ is uniquely determined.
Thus we see that $\mu(\tilde{\mathcal{D}})\simeq{\rm Pic}^0(\Gamma)$ as a group.

Hereafter, we fix $D^\ast$ as follows
\begin{align*}
D^\ast
=
\begin{cases}
\displaystyle\frac{g}{2}(V_0+V_0^\prime)&\mbox{for even $g$,}\\
\displaystyle\frac{g-1}{2}(V_0+V_0^\prime)+V_0&\mbox{for odd $g$.}\\
\end{cases}
\end{align*}
Define the element $o\in{\rm Sym}^g(\Gamma)$ to be
\begin{align*}
o
:=
\begin{cases}
\DIS\bigcup_{i=1}^{g/2}\left\{V_{2i-1},V_{2i-1}^\prime\right\}&\mbox{for even $g$},\\
\{V_0\}\cup\DIS\left(\bigcup_{i=1}^{g-1/2}\left\{V_{2i},V_{2i}^\prime\right\}\right)&\mbox{for odd $g$.}\\
\end{cases}
\end{align*}
Then $o$ is the unit of addition of the group $\mu(\tilde{\mathcal{D}})$ \cite{Nobe12,Nobe13}.

\section{Tropical eigenvector maps}\label{sec:TropicalEigenvectorMap}
Let  the phase space of the UD-pTL be
\begin{align*}
\mathcal{T}:=\left\{(\bJ,\bW)\ |\ \sum_{i=1}^{g+1}J_i<\sum_{i=1}^{g+1}W_i\right\}
\end{align*}
and the moduli space of $\Gamma$ be $\mathcal{C}:=\left\{(C_{-1},C_0,\ldots,C_g)\right\}$.
Consider the map $\psi:\mathcal{T}\to\mathcal{C}$ defined by (\ref{eq:CJW}) and set 
\begin{align*}
\mathcal{T}_C
:=
\psi^{-1}(C_{-1},C_0,\ldots,C_g)\subset\mathcal{T}.
\end{align*}
The set $\mathcal{T}_C$ is called the isospectral set of the UD-pTL.

Now remember the periodic discrete Toda lattice (pdTL) $\chi:\mathbb{C}^{2g+2}\to\mathbb{C}^{2g+2}; (\bI,\bV)$ $\mapsto(\bbI,\bbV)$ given by
\begin{align}
\bar I_i+\bar V_{i-1}=I_i+V_i,
\qquad
\bar V_i\bar I_i=I_{i+1}V_i,
\label{eq:pdTL}
\end{align}
where we put $\bI=(I_1,\ldots,I_{g+1})$ and $\bV=(V_1,\ldots,V_{g+1})$  \cite{HTI93}.
By applying the procedure of ultradiscretization to the pdTL, we obtain the UD-pTL \cite{YT02, KT02}.
Let $\varphi(x,y)=\left(\varphi_1,\cdots,\varphi_g,-\varphi_{g+1}\right)^T$ be the eigenvector of the Lax matrix
\begin{align*}
L
:=
&\left[
\begin{matrix}
I_2+V_1&1&&(-1)^gI_1V_1/y\cr
I_2V_2&\ddots&\ddots&\cr
&\ddots&\ddots&1\cr
(-1)^gy&&I_{g+1}V_{g+1}&I_1+V_{g+1}\cr
\end{matrix}
\right]
\end{align*}
of the pdTL.
Here $-x$ is the eigenvalue of $L$ and $y$ is the spectral parameter.

Let $\tilde f$ be the polynomial $y|x\mathbb{I}+L|$ in $x$ and $y$.
Then $\tilde f$ has the form
\begin{align*}
\tilde f(x,y)
=
y^2
+
y\left(
x^{g+1}+c_gx^g+\cdots+c_1x+c_0
\right)
+
c_{-1}
\end{align*}
and defines the spectral curve $\tilde\gamma$ of the pdTL, which is a hyperelliptic curve of genus $g$ for appropriate choice of $c_i$'s \cite{IT08}.
Note that the coefficients $c_{-1},c_0,\ldots,c_g$ are subtraction-free polynomials in $\bI$ and $\bV$ and are the conserved quantities of the pdTL.

Consider the eigenvalue equation of $L$
\begin{align}
\left(
x\mathbb{I}+L
\right)
\varphi(x,y)
=
0.
\label{eq:sse}
\end{align}
Let the $(i,j)$-entry of $L$ be $l_{ij}$.
By applying the Cramer formula, each element of $\varphi(x,y)$ is explicitly given by
\begin{align*}
\varphi_i(x,y)
=
\left|
\begin{matrix}
&l_{11}+x&\cdots&l_{1,i-1}&l_{1,g+1}&l_{1,i+1}&\cdots&l_{1g}\cr
&\vdots&&\vdots&\vdots&\vdots&&\vdots\cr
&l_{g1}&\cdots&l_{g,i-1}&l_{g,g+1}&l_{g,i+1}&\cdots&l_{gg}+x\cr
\end{matrix}
\right|
\end{align*}
for $i=1,2,\ldots,g$ and
\begin{align*}
\varphi_{g+1}(x)
=
\left|x\mathbb{I}+L_{\bar\Lambda}\right|,
\end{align*}
where $L_{\bar\Lambda}=(l_{ij})_{1\leq i,j\leq g}$.

Let $\mathcal{U}_c$ be the isospectral set of the pdTL.
The eigenvector map $\phi$ of the pdTL is defined to be a map $\phi:\mathcal{U}_c\to{\rm Pic}^g\mkern2mu(\gamma):=\mathcal{D}_g(\gamma)/\mathcal{D}_l(\gamma);$
\begin{align*}
(\bI,\bV)\mapsto\phi(\bI,\bV)\equiv P_1+\cdots+P_g
\quad
\mbox{(mod $\mathcal{D}_l(\gamma)$)},
\end{align*}
where $P_1,P_2, \cdots,P_g$ are points on $\tilde\gamma$ such that they are the common $g$ zeros of the rational functions $\varphi_1(x,y),\varphi_2(x,y),\ldots,\varphi_g(x,y)$ in the eigenvector $\varphi(x,y)$ \cite{vMM79,Iwao08}.

We can easily see that the first two entries $\varphi_1(x,y)$ and $\varphi_2(x,y)$ of $\varphi(x,y)$ have exactly $g$ zeros in common, and these $g$ zeros define $\phi$.
Expand $\varphi_1(x,y)$ with respect to the first column:
\begin{align}
\varphi_1(x,y)
&=
\frac{(-1)^g}{y}I_1V_1
\left|
\begin{matrix}
I_3+V_2+x&1&&\cr
I_3V_3&\ddots&\ddots&\cr
&\ddots&\ddots&1\cr
&&I_gV_g&I_{g+1}+V_g+x\cr
\end{matrix}
\right|
-
(-1)^{g}.
\label{eq:varphi1}
\end{align}
Also expand $\varphi_2(x,y)$ with respect to the second column:
\begin{align}
&\varphi_2(x,y)
=
(-1)^g\left(I_2+V_1+x\right)
-
\frac{(-1)^{g}I_1I_2V_1V_2}{y}
\left|
\begin{matrix}
I_4+V_3+x&1&&\cr
I_4V_4&\ddots&\ddots&\cr
&\ddots&\ddots&1\cr
&&I_{g}V_{g}&I_{g+1}+V_{g}+x\cr
\end{matrix}
\right|.
\label{eq:varphi2}
\end{align}

Now consider the following formula \cite{Nobe13}:
\begin{align}
\left|
\begin{matrix}
I_2+V_1+x&1&&\cr
I_2V_2&\ddots&\ddots&\cr
&\ddots&\ddots&1\cr
&&I_{g}V_{g}&I_{g+1}+V_{g}+x\cr
\end{matrix}
\right|
&=
\frac{
\left|x\mathbb{I}+L(\bI_{2,{g+1}};\bV_{1,{g}})\right|-y}{x}
\label{eq:formulavarphigp1}\\
&=
x^g
+
\sum_{i=1}^gc_{i}(\bI_{2,{g+1}};\bV_{1,{g}})x^{i-1},
\nn
\end{align}
where we put
\begin{align*}
&\bI_{i,j}:=(\underset{i-1}{\underbrace{0,\ldots,0}},I_i,I_{i+1},\ldots,I_j,\underset{g+1-j}{\underbrace{0,\ldots,0}}),
\\
&\bV_{i,j}:=(\underset{i-1}{\underbrace{0,\ldots,0}},V_i,V_{i+1},\ldots,V_j,\underset{g+1-j}{\underbrace{0,\ldots,0}})
\end{align*}
for $i\leq j$, $i,j=1,2,\ldots,g+1$.
It should be noted that \eqref{eq:formulavarphigp1} is the subtraction-free polynomial $\varphi_{g+1}(x)$ in $x$, the $(g+1)$-th entry of the eigenvector $\varphi(x,y)$.

Setting $I_{2}=V_1=0$ in \eqref{eq:formulavarphigp1}, we have
\begin{align*}
&\left|
\begin{matrix}
I_3+V_2+x&1&\cr
I_3V_3&\ddots&\ddots&\cr
&\ddots&\ddots&1\cr
&&I_{g}V_{g}&I_{g+1}+V_{g}+x\cr
\end{matrix}
\right|
=
x^{g-1}
+
\sum_{i=2}^gc_{i}(\bI_{3,{g+1}};\bV_{2,{g}})x^{i-2},
\end{align*}
where we use the fact $c_1(\bI_{3,{g+1}};\bV_{2,{g}})=0$.
Moreover setting $I_{3}=V_2=0$, we obtain
\begin{align*}
&\left|
\begin{matrix}
I_4+V_3+x&1&\cr
I_4V_4&\ddots&\ddots&\cr
&\ddots&\ddots&1\cr
&&I_{g}V_{g}&I_{g+1}+V_{g}+x\cr
\end{matrix}
\right|
=
x^{g-2}
+
\sum_{i=3}^gc_{i}(\bI_{4,{g+1}};\bV_{3,{g}})x^{i-3},
\end{align*}
where we use the fact $c_2(\bI_{4,{g+1}};\bV_{3,{g}})=0$.
Thus we find the following explicit formulae for $\varphi_1$ and $\varphi_2$
\begin{align*}
\varphi_1(x,y)
&=
\frac{(-1)^g}{y}I_1V_1
\left\{
x^{g-1}
+
\sum_{i=2}^gc_{i}(\bI_{3,{g+1}};\bV_{2,{g}})x^{i-2}
\right\}
-
(-1)^{g},\\
\varphi_2(x,y)
&=
(-1)^g\left(I_2+V_1+x\right)
-
\frac{(-1)^{g}}{y}I_1I_2V_1V_2
\left\{
x^{g-2}
+
\sum_{i=3}^gc_{i}(\bI_{4,{g+1}};\bV_{3,{g}})x^{i-3}
\right\}.
\end{align*}

From the first equation in \eqref{eq:sse} we have
\begin{align}
\varphi_{g+1}(x)
=
\frac{(-1)^gy}{I_1V_1}
\left\{
\left(I_1+V_1+x\right)\varphi_1(x,y)
+
\varphi_2(x,y)
\right\}.
\label{eq:phig12}
\end{align}
The remaining equations in \eqref{eq:sse} imply
\begin{align}
&I_{i+1}V_{i+1}\varphi_{i}+\left(I_{i+2}+V_{i+1}+x\right)\varphi_{i+1}+\varphi_{i+2}
=
0
\label{eq:phirel}
\end{align}
for $i=1,2,\ldots,g-2$ and
\begin{align}
&(-1)^gy\varphi_1(x,y)+I_{g+1}V_{g+1}\varphi_{g}(x,y)-\left(I_1+V_{g+1}\right)\varphi_{g+1}(x)
=
0.
\label{eq:phitildef}
\end{align}
Note that \eqref{eq:phitildef} is equivalent to the defining polynomial $\tilde f(x,y)$ of $\tilde\gamma$.

Now we apply the procedure of ultradiscretization to $\varphi_1$ and $\varphi_2$.
Replace $x$ and $-y$ with $e^{-X/\epsilon}$ and $e^{-Y/\epsilon}$, respectively.
Also replace $I_i$ and $V_i$ with $e^{-J_i/\epsilon}$ and $e^{-W_i/\epsilon}$ for $i=1,2,\ldots,g-1$, respectively.
Then the polynomials $(-1)^{g-1}y\varphi_1(x,-y)$ and $(-1)^gy\varphi_2(x,-y)$ are subtraction-free and we obtain tropical polynomials
\begin{align*}
G_1(X,Y)
:=&
-\lim_{\epsilon\to0}\epsilon\log\left(\frac{(-1)^{g-1}y}{I_1V_1}\varphi_1(x,-y)\right)\\
=&
\left\lfloor
\underset{2\leq i\leq g}{\left\lfloor C_{i}(\bJ_{3,{g+1}};\bW_{2,{g}})+(i-2)X\right\rfloor},
(g-1)X,
Y-J_1-W_1
\right\rfloor
\end{align*}
and
\begin{align*}
&G_2(X,Y)
:=
-\lim_{\epsilon\to0}\epsilon\log\left(\frac{(-1)^gy}{I_1I_2V_1V_2}\varphi_2(x,-y)\right)\\
&=
\left\lfloor
\underset{3\leq i\leq g}{\left\lfloor C_{i}(\bJ_{4,{g+1}};\bW_{3,{g}})+(i-3)X\right\rfloor},
(g-2)X,
Y
+
\left\lfloor
J_2,W_1,X
\right\rfloor
-
\sum_{i=1}^2\left(J_i+W_i\right)
\right\rfloor
\end{align*}
in the limit $\epsilon\to0$.
Here $C_i$ ($i=1,2,\ldots,g-1)$ is the coefficient in the defining polynomial $F$ of $\Gamma$ and we put
\begin{align*}
&\bJ_{i,j}:=(\underset{i-1}{\underbrace{\infty,\ldots,\infty}},J_i,J_{i+1},\ldots,J_j,\underset{g+1-j}{\underbrace{\infty,\ldots,\infty}}),
\\
&\bW_{i,j}:=(\underset{i-1}{\underbrace{\infty,\ldots,\infty}},W_i,W_{i+1},\ldots,W_j,\underset{g+1-j}{\underbrace{\infty,\ldots,\infty}})
\end{align*}
for $i\leq j$, $i,j=1,2,\ldots,g+1$.

Define tropical plane curves $L_1$ and $L_2$ by using these tropical polynomials:
\begin{align*}
&L_1
:=
\left\{
P\in\R^2\ |\ 
\mbox{$G_1$ is not differentiable at $P$}
\right\},\\
&L_2
:=
\left\{
P\in\R^2\ |\ 
\mbox{$G_2$ is not differentiable at $P$}
\right\}.
\end{align*}
We may use the notation $G_1^t(X,Y), G_2^t(X,Y)$ and $L_1^t, L_2^t$ for $\bJ^t$ and $\bW^t$ for $t=0,1,\ldots$.

By embedding these tropical curves into tropical projective plane $\mathbb{TP}^2$, we can prove that $L_1$ intersects $L_2$ exactly at $g$ points $P_1,P_2,\ldots,P_g$.  
Moreover, by \eqref{eq:phig12}--\eqref{eq:phitildef}, we find that the $g$ intersection points $P_1,P_2,\ldots,P_g$ of $L_1$ and $L_2$ are on $\Gamma$.

We define the tropical eigenvector map $\phi:\mathcal{T}_C\to{\rm Pic}^g(\Gamma):=\mathcal{D}_g^+(\Gamma)/\mathcal{D}_l(\Gamma)$ to be
\begin{align*}
\phi(\bJ,\bW)
\equiv
P_1+P_2+\cdots+P_g
\quad
(\mbox{mod $\mathcal{D}_l(\Gamma)$}),
\end{align*}
where $P_1,P_2,\ldots,P_g$ are the intersection points of $L_1$, $L_2$ and $\Gamma$.

\section{A geometric realization of UD-pTL}
\label{sec:AGROP}
Define $T\in\mathcal{D}_g^+(\Gamma)$ to be
\begin{align}
T
:=
\begin{cases}
\DIS
V_{0}
+
V_g^\prime
+
\sum_{i=1}^{(g-2)/2}\left(V_{2i}+V_{2i}^\prime\right)
&
\mbox{for even $g$,}\\
\DIS
B
+
V_1
+
V_g^\prime
+
\sum_{i=2}^{(g-1)/2}\left(V_{2i-1}+V_{2i-1}^\prime\right)
&
\mbox{for odd $g$}\\
\end{cases}
\label{eq:tepBBS}
\end{align}
for $g\geq2$.
Here $B=(C_g,C_{-1}-C_{g-1}-(g-1)C_g)\in\overline{V_0V_0^\prime}\subset\alpha_1$ is the unique point such that $B+V_1-2V_0$ is the principal divisor of a rational function on $\Gamma$. 
For $g=1$, we define $T$ to be the point $(C_1,C_0)\in\tilde{\mathcal{D}}$ on the edge $\overline{V_0V_0^\prime}$.

We easily observe that $T\in\tilde{\mathcal{D}}$.
Moreover, we obtain the following theorem concerning time evolution of the UD-pTL and an addition on ${\rm Sym}^g(\Gamma)$.
\begin{thm}[\cite{Nobe13}]\label{thm:tepBBSonSymg}
Set $\tau=\mu(T)\in {\rm Sym}^g(\Gamma)$.
Then, for any $g$, the following diagram is commutative
\begin{align*}
\begin{CD}
\mathcal{T}_C @> \mu\circ\phi >>{\rm Sym}^g(\Gamma)\\
@V (\ref{eq:pbbs}) VV @VV \oplus\tau V\\
\mathcal{T}_C @>> \mu\circ\phi > {\rm Sym}^g(\Gamma).\\
\end{CD}
\end{align*}
\end{thm}

Let $P_1,P_2,\ldots,P_g$ be the points on $\Gamma$ given by the tropical eigenvector map $\phi$.
Also let $d_P=\mu(D_P)$ for $D_P=P_1+P_2+\cdots+P_g$.
Then the addition of points on ${\rm Sym}^g(\Gamma)$ in theorem \ref{thm:tepBBSonSymg} is written as
\begin{align}
d_{\bar P}
=
d_P\oplus\tau
\quad
\Longleftrightarrow
\quad
\begin{cases}
d_{Q}
\oplus
d_P
\oplus
\tau
=
o,\\
d_{Q}
\oplus
d_{\bar P}
=
o.\\
\end{cases}
\label{eq:addtropsym}
\end{align}
Since $\phi(T)\equiv V_g^\prime-V_0^\prime$ (mod $\mathcal{D}_l(\Gamma)$) for any $g$, (\ref{eq:addtropsym}) is written by the divisors on ${\rm Pic}^0(\Gamma)$:
\begin{align}
&D_Q+D_P+V_g^\prime-V_0^\prime-2D^\ast
\equiv
0
\quad
\mbox{(mod $\mathcal{D}_l(\Gamma)$)},
\label{eq:addtroppic}\\
&D_Q
+
D_{\bar P}
-
2D^\ast
\equiv
0
\quad
\mbox{(mod $\mathcal{D}_l(\Gamma)$)},
\label{eq:addtroppic2}
\end{align}
where $D_Q=Q_1+\cdots+Q_g=\mu^{-1}(d_{Q})$.
Hereafter, we assume that the divisor $D_P=P_1+\cdots+P_g$ is in $\tilde{\mathcal{D}}$.

Let $M$ be a number such that $M\geq\left\lceil\bJ,\bW\right\rceil$.
Define $S_jC_i$ and $T_kC_i$ for $i=0,\ldots,g$ and $j,k=1,2,\ldots,g+1$ to be
\begin{align*}
&S_jC_i
:=
C_i(J_1,\ldots,J_j-(g-i+1)M,\ldots,J_{g+1};\bW)+(g-i+1)M,\\
&T_kC_i
:=
C_i(\bJ;W_1,\ldots,W_{k-1},W_k+(g-i+1)M,W_{k+1},\ldots,W_{g+1}).
\end{align*}
Note that $S_jC_i$ eliminates the terms in $C_i$ not containing $J_j$ and $T_kC_i$ the terms in $C_i$ containing $W_k$.

Define tropical polynomials
\begin{align}
&H_1(X,Y)
:=
\begin{cases}
\left\lfloor
\underset{0\leq i\leq g}{\left\lfloor S_1C_i+iX\right\rfloor},Y
\right\rfloor
&\mbox{for even $g$,}\\[10pt]
\left\lfloor
\underset{0\leq i\leq g}{\left\lfloor T_{g+1}C_i+iX\right\rfloor},(g+1)X,Y
\right\rfloor
&\mbox{for odd $g$}
\end{cases}
\label{eq:trophk}
\end{align}
and
\begin{align*}
H_2(X,Y)
:=
\left\lfloor
\acute C_{0},
\acute C_{1}+X,
\ldots,
\acute C_{g-1}+(g-1)X,
C_g+gX,
(g+1)X,
Y
\right\rfloor,
\end{align*}
where $\acute C_i:=\left\lfloor C_i, S_1C_i, J_1+C_{i+1}(\bJ_{2,g+1};\bW_{2,g+1})\right\rfloor$ for $i=0,1,\ldots,g-1$.

By using $H_1$ and $H_2$, we define the tropical curves $K_1$ and $K_2$, respectively
\begin{align*}
&K_1
:=
\left\{
P\in\R^2\ |\ 
\mbox{$H_1$ is not differentiable at $P$}
\right\},
\\
&K_2
:=
\left\{
P\in\R^2\ |\ 
\mbox{$H_2$ is not differentiable at $P$}
\right\}.
\end{align*}
These tropical plane curves are tropical analogues of the plane curves which intersect the spectral curve $\tilde\gamma$ of the pdTL and give the time evolution of the pdTL \cite{Nobe13}.
We may use the notation $H_1^t(X,Y), H_2^t(X,Y)$ and $K_1^t, K_2^t$ for $\bJ^t$ and $\bW^t$ for $t=0,1,\ldots$.

Denote the restriction of the rational function $H_1$ on $\Gamma$ by $H_1|\Gamma$.
We then have the following proposition.
\begin{prp}[\cite{Nobe13}]\label{prop:troph1}
Assume that the points $P_1,P_2,\ldots,P_g\in\Gamma$ are on the tropical curve $K_1$.
Then the rational function $H_1|\Gamma$ on $\Gamma$ satisfies
\begin{align}
(H_1|\Gamma)
=
D_Q+D_P+V_g^\prime-V_0^\prime-2D^\ast.
\label{eq:Hpd}
\end{align}
\end{prp}

Thus we see that the addition \eqref{eq:addtroppic} is realized by using the intersection of $\Gamma$ and $K_1$ defined by $H_1$, and hence $Q_1,\ldots,Q_g$ are the intersection points of $\Gamma$ and $K_1$.
Moreover, if $g$ is an even number we have (see \cite{Nobe12})
\begin{align*}
D_Q
+
D_{Q^\prime}
-
2D^\ast
\equiv
0
\quad
\mbox{(mod $\mathcal{D}_l(\Gamma)$)}
\end{align*}
for $Q_1,\ldots,Q_g\in\Gamma$.
This implies $D_{\bar P}=D_{Q^\prime}$.
Thus the time evolution of the UD-pTL is realized by using the intersection of tropical plane curves $\Gamma$ and $K_1$ if $g$ is an even number.

Now assume that $g$ is an odd number.
The restriction $H_2|\Gamma$ of rational function $H_2$ on $\Gamma$ has poles of oder $g$ at $V_0$ and $V_0^\prime$.
This implies
\begin{align*}
\left(
H_2|\Gamma
\right)
=
D_R+D_{R^\prime}+V_0-V_0^\prime-2D^\ast,
\end{align*}
where $R_1,\ldots,R_g$ and $R_1^\prime,\ldots,R_g^\prime$ are the intersection points of $\Gamma$ and $K_2$ defined by $H_2$.
There uniquely exists a point $Z_0$ on $\alpha_1\setminus\alpha_{1,2}$ such that 
\begin{align*}
V_0^\prime+Z_0-V_0-R_1^\prime
\equiv
0
\quad
\mbox{(mod $\mathcal{D}_l(\Gamma)$)}.
\end{align*}
We then find
\begin{align*}
(H_2|\Gamma)
\equiv
D_R+Z_0+R_2^\prime+\cdots+R_g^\prime-2D^\ast
\quad
\mbox{(mod $\mathcal{D}_l(\Gamma)$)}.
\end{align*}

Since there exists at least one intersection point of $\Gamma$ and $K_1$ on the upper half of $\alpha_1\setminus\alpha_{1,2}$, we can assume that $Q_1$ is the one.
Then there exist rational functions $G_1$ and $G_2$ on $\Gamma$ such that
\begin{align*}
\left(G_1\right)
&=
Z_1+Q_1-Z_0-R_1,\\
\left(G_2\right)
&=
Q_2+\cdots+Q_g+Q_2^\prime+\cdots+Q_g^\prime-R_2-\cdots-R_g-R_2^\prime-\cdots-R_g^\prime.
\end{align*}
Then we have
\begin{align*}
(G_1+G_2)
&=
D_Q
+
Z_1
+
Q_2^\prime
+
\cdots
+
Q_g^\prime
-
Z_0
-
D_R
-
R_2^\prime-\cdots-R_g^\prime.
\end{align*}
It immediately follows
\begin{align}
\left(H_2|\Gamma\right)
\equiv
D_Q
+
Z_1
+
Q_2^\prime
+
\cdots
+
Q_g^\prime
-
2D^\ast
\equiv
0
\quad
(\mbox{mod $\mathcal{D}_l(\Gamma)$}).
\label{eq:H2pd}
\end{align}

Comparing \eqref{eq:addtroppic2} with \eqref{eq:H2pd}, we find
\begin{align*}
D_{\bar P}
=
Z_1+Q_2^\prime+\cdots+Q_g^\prime.
\end{align*}
Remember that $Q_1,\ldots,Q_g$ are the intersection points of $\Gamma$ and $K_1$, and $Z_1$ is the unique point on $\Gamma$ determined by the intersection points $R_1,R_1^\prime$ of $\Gamma$ and $K_2$ by the formula
\begin{align*}
&Z_1+Q_1+V_0^\prime-R_1-R_1^\prime-V_0
\equiv
0
\quad
(\mbox{mod $\mathcal{D}_l(\Gamma)$}).
\end{align*}
Thus the time evolution of the UD-pTL is realized by using the intersection of the tropical plane curves $\Gamma$, $K_1$ and $K_2$ if $g$ is an odd number.
We obtain the following theorem.

\begin{thm}[\cite{Nobe13}]
Let $D_P$ and $D_Q$ be elements of $\tilde{\mathcal{D}}$ satisfying (\ref{eq:addtroppic}).
Also let $T$ be the element of $\tilde{\mathcal{D}}$ given by (\ref{eq:tepBBS}).
Put $d_{P}=\mu(D_P)\in{\rm Sym}^g(\Gamma)$ and $\tau=\mu(T)\in{\rm Sym}^g(\Gamma)$.
Then the element $d_{\bar P}\in{\rm Sym}^g(\Gamma)$ defined by the addition 
\begin{align*}
d_{\bar P}
=
d_P\oplus\tau
\end{align*}
is explicitly given by the formula
\begin{align*}
d_{\bar P}
=
\begin{cases}
\left\{
Q_1^\prime,Q_2^\prime,\ldots,Q_g^\prime
\right\}
&\mbox{for even $g$},\\
\left\{
Z_1,Q_2^\prime,\ldots,Q_g^\prime
\right\}
&\mbox{for odd $g$.}\\
\end{cases}
\end{align*}
\end{thm}

In appendix \ref{sec:example}, we give an explicit computation of the time evolution of the pBBS imposing a certain initial condition on the integral lattice via tropical curve intersection. 

\section{Conclusion}
We establish a geometric realization of the UD-pTL via the tropical curve intersection of its spectral curve $\Gamma$ and two plane curves $K_1$ and $K_2$.
Namely, the linear flow on the tropical Jacobian of $\Gamma$ equivalent to the time evolution of the UD-pTL is translated into the intersection of tropical curves $\Gamma$, $K_1$ and $K_2$.
The rational functions $F$, $H_1$ and $H_2$ which respectively define $\Gamma$, $K_1$ and $K_2$ are explicitly given by using the conserved quantities $C_{-1},C_0,\ldots,C_g$ of the UD-pTL.
Moreover, the tropical eigenvector map, which maps a point in the isospectral set $\mathcal{T}_C$ of the UD-pTL into a set of points on $\Gamma$, is explicitly given by using two tropical plane curves $L_1$ and $L_2$.
The rational functions $G_1$ and $G_2$ which respectively define these tropical plane curves are also given by using the conserved quantities of the UD-pTL.
Thus, if initial values of the UD-pTL are given then we can completely realize the time evolution of the UD-pTL via tropical plane curves $\Gamma$, $L_1$, $L_2$, $K_1$ and $K_2$.
We observe that the points $P_1^t,P_2^t,\ldots,P_g^t$ on $\Gamma$ corresponding to the variables $\bJ^t, \bW^t$ of the UD-pTL move the cycles $\alpha_1,\alpha_2,\ldots,\alpha_g$ in $\Gamma$ in a counterclockwise direction as $t$ increasing, respectively.

The pBBS studied in this article is a fundamental ultradiscrete integrable system and is associated with the crystal basis of the quantum group ${U_q(\widehat{\mathfrak{sl}}_2)}$ \cite{HHIKTT01}.
To establish geometric realizations for members in a wide class of ultradiscrete integrable systems such as box-ball systems associated with the crystal bases of the quantum groups ${U_q(\hat{\mathfrak{g}})}$ of affine Lie algebras $\hat{\mathfrak{g}}$ other than $\widehat{\mathfrak{sl}}_2$ is a further problem.

\appendix
\section{An example of geometric realization of pBBS}\label{sec:example}
In this section, we show explicit computation of the tropical geometric realization of the UD-pTL discussed above.
In particular, we consider the pBBS case, that is, we take the initial values in positive integers.
Throughout this section, we fix $g=2$.

\begin{figure}[htbp]
\centering
{\unitlength=.03in{\def\arraystretch{1.0}
\begin{picture}(150,75)(0,-62)
\thicklines
\thicklines
\multiput(0,0)(10,0){16}{\line(0,1){10}}
\put(0,0){\line(1,0){150}}
\put(0,10){\line(1,0){150}}

\multiput(5,5)(10,0){3}{\circle{8}}
\multiput(65,5)(10,0){2}{\circle{8}}
\multiput(105,5)(10,0){1}{\circle{8}}
\multiput(0,-15)(10,0){16}{\line(0,1){10}}
\put(0,-15){\line(1,0){150}}
\put(0,-5){\line(1,0){150}}

\multiput(35,-10)(10,0){3}{\circle{8}}
\multiput(85,-10)(10,0){2}{\circle{8}}
\multiput(115,-10)(10,0){1}{\circle{8}}
\multiput(0,-30)(10,0){16}{\line(0,1){10}}
\put(0,-30){\line(1,0){150}}
\put(0,-20){\line(1,0){150}}

\multiput(65,-25)(10,0){2}{\circle{8}}
\multiput(105,-25)(10,0){1}{\circle{8}}
\multiput(125,-25)(10,0){3}{\circle{8}}
\multiput(0,-45)(10,0){16}{\line(0,1){10}}
\put(0,-45){\line(1,0){150}}
\put(0,-35){\line(1,0){150}}

\multiput(85,-40)(10,0){2}{\circle{8}}
\multiput(115,-40)(10,0){1}{\circle{8}}
\multiput(5,-40)(10,0){3}{\circle{8}}
\multiput(0,-60)(10,0){16}{\line(0,1){10}}
\put(0,-60){\line(1,0){150}}
\put(0,-50){\line(1,0){150}}

\multiput(105,-55)(10,0){1}{\circle{8}}
\multiput(125,-55)(10,0){2}{\circle{8}}
\multiput(35,-55)(10,0){3}{\circle{8}}


\end{picture}
}}
\caption{Time evolution of the pBBS for the initial values \eqref{eq:ivpBBS} from $t=0$ to $4$.
}
\label{fig:pBBSex}
\end{figure}
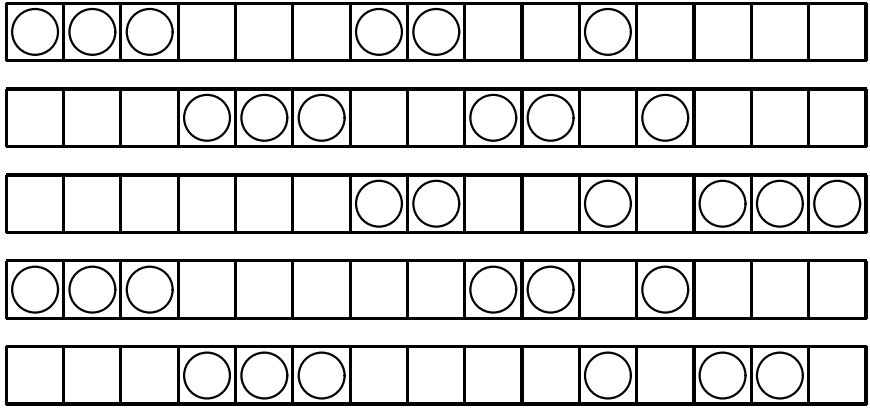

Let us consider the following initial values of the pBBS
\begin{align}
\bJ^0=(J_1^0,J_2^0,J_3^0)=(3,2,1),\quad
\bW^0=(W_1^0,W_2^0,W_3^0)=(3,2,4).
\label{eq:ivpBBS}
\end{align}
The time evolution of the pBBS imposing this initial condition from $t=1$ to $4$ is computed as follows
\begin{align*}
&
\bJ^1=(J_1^1,J_2^1,J_3^1)=(3,2,1),\quad
\bW^1=(W_1^1,W_2^1,W_3^1)=(2,1,6),
\\
&
\bJ^2=(J_1^2,J_2^2,J_3^2)=(2,1,3),\quad
\bW^2=(W_1^2,W_2^2,W_3^2)=(2,1,6),
\\
&
\bJ^3=(J_1^3,J_2^3,J_3^3)=(2,1,3),\quad
\bW^3=(W_1^3,W_2^3,W_3^3)=(1,3,5),
\\
&
\bJ^4=(J_1^4,J_2^4,J_3^4)=(1,2,3),\quad
\bW^4=(W_1^4,W_2^4,W_3^4)=(1,4,4).
\end{align*}
The realization of the pBBS with boxes and balls is illustrated in figure \ref{fig:pBBSex}.

The spectral curve $\Gamma$ of the pBBS is given by 
\begin{align*}
F(X,Y)
=
\left\lfloor
2X,
Y+\left\lfloor
3X,C_2+2X,C_1+X,C_0
\right\rfloor,
C_{-1}
\right\rfloor,
\end{align*}
where
\begin{align*}
&C_2=\left\lfloor
J_1,J_2,J_3,W_1,W_2,W_3
\right\rfloor,
\\
&C_1=\left\lfloor
J_1+J_2,J_2+J_3,J_3+J_1,W_1+W_2,W_2+W_3,\right.
\\
&\qquad\qquad\qquad\left.
W_3+W_1,J_1+W_2,J_2+W_3,J_3+W_1
\right\rfloor,
\\
&C_0
=
J_1+J_2+J_3,
\\
&C_{-1}=J_1+J_2+J_3+W_1+W_2+W_3.
\end{align*}

Substituting the initial values \eqref{eq:ivpBBS} into the coefficients $C_{-1},C_0,C_1,C_2$, we obtain
\begin{align*}
&C_2
=
C_2^0
=
\left\lfloor
3,2,1,3,2,4
\right\rfloor
=
1,
\\
&C_1
=
C_1^0
=
\left\lfloor
3+2,2+1,1+3,3+2,2+4,
4+3,3+2,2+4,1+3
\right\rfloor
=
3,
\\
&C_0
=
C_0^0
=
3+2+1
=
6,
\\
&C_{-1}
=
C_{-1}^0
=
3+2+1+3+2+4
=
15
\end{align*}
and hence
\begin{align}
F(X,Y)
=
\left\lfloor
2X,
Y+\left\lfloor
3X,1+2X,3+X,6
\right\rfloor,
15
\right\rfloor.
\label{eq:Fexample}
\end{align}
Figure \ref{fig:evmt0} shows the spectral curve $\tilde\Gamma$ given by \eqref{eq:Fexample}, which is a tropical hyperelliptic curve of genus 2.
The vertices in $\tilde\Gamma$ are as follows
\begin{align*}
V_0=(1,12),\quad
V_1=(2,10),\quad
V_2=(3,9),\quad
V_0^\prime=(1,3),\quad
V_1^\prime=(2,5),\quad
V_2^\prime=(3,6).
\end{align*}

The tropical plane curves $L_1$ and $L_2$ which define the tropical eigenvector map are given by the tropical polynomials
\begin{align*}
G_1(X,Y)
&=
\left\lfloor
{\left\lfloor C_{2}(\bJ_{3,{3}};\bW_{2,{2}})\right\rfloor},
X,
Y-J_1-W_1
\right\rfloor\\
&=\left\lfloor
J_3,W_2,
X,
Y-J_1-W_1
\right\rfloor,
\\
G_2(X,Y)
&=
\left\lfloor
0,
Y
+
\left\lfloor
J_2,W_1,X
\right\rfloor
-
\sum_{i=1}^2\left(J_i+W_i\right)
\right\rfloor,
\end{align*}
where we use 
\begin{align*}
&C_{2}(\bJ_{3,{3}};\bW_{2,{2}})
=
C_{2}(\infty,\infty,J_3;\infty,W_2,\infty)
=
\left\lfloor
J_3,W_2
\right\rfloor.
\end{align*}
For the initial values \eqref{eq:ivpBBS}, we obtain
\begin{align*}
G_1^0(X,Y)
&=\left\lfloor
1,2,
X,
Y-6
\right\rfloor,
\\
G_2^0(X,Y)
&=
\left\lfloor
0,
Y
+
\left\lfloor
2,3,X
\right\rfloor
-
10
\right\rfloor
=
\left\lfloor
0,
Y-8,
Y+X-10
\right\rfloor.
\end{align*}
Figure \ref{fig:evmt0} shows $\Gamma$, $L_1^0$ and $L_2^0$ and their intersection points $P_1^0=(1,9)$ and $P_2^0=(2,7)$.
By applying the eigenvector map repeatedly, we obtain the sequence 
\begin{align*}
&P_1^0=(1,9),&&P_1^1=(1,7),&&P_1^2=(1,5),&&P_1^3=(1,4),&&P_1^4=(1,3),&&\ldots&\\
&P_2^0=(2,7),&&P_2^1=(2,6),&&P_2^2=(2,5),&&P_2^3=(3,6),&&P_2^4=(3,7),&&\ldots&
\end{align*}
of points on $\Gamma$ corresponding to the sequence $\bJ^0,\bW^0,\bJ^1,\bW^1,\ldots$ (see figure \ref{fig:k1total}).
We observe that the points $P_1^t$ and $P_2^t$ move the cycles $\alpha_1$ and $\alpha_2$ in a contourclockwise direction as $t$ increasing, respectively.

\begin{figure}[t]
\centering
\subfigure[Eigenvector map for $t=0$]
{\unitlength=.3in{\def\arraystretch{1.0}
\begin{picture}(5,11)(0,2)
\put(1,3){\line(1,2){1}}
\put(1,3){\line(0,1){9}}
\put(1,12){\line(1,-2){1}}
\put(2,5){\line(0,1){5}}
\put(2,5){\line(1,1){1}}
\put(2,10){\line(1,-1){1}}
\put(3,6){\line(0,1){3}}
\put(1,3){\line(-1,-3){.3}}
\put(1,12){\line(-1,3){.3}}
\put(3,6){\line(1,0){3}}
\put(3,9){\line(1,0){3}}
\thicklines
\put(1,7){{\line(1,0){5}}}
\put(1,7){{\line(0,1){6}}}
\put(1,7){{\line(-1,-1){1}}}
\dashline{.2}(2,8)(6,8)
\dashline{.2}(2,8)(2,2)
\dashline{.2}(2,8)(0,10)
\put(2,7){\circle{.3}}
\put(1,9){\circle{.3}}
\put(2,7){\circle*{.1}}
\put(1,9){\circle*{.1}}
\put(.6,3.0){\makebox(0,0){$V_0^\prime$}}
\put(2.3,4.7){\makebox(0,0){$V_1^\prime$}}
\put(3.3,5.7){\makebox(0,0){$V_2^\prime$}}
\put(.6,11.9){\makebox(0,0){$V_0$}}
\put(2.3,10.3){\makebox(0,0){$V_1$}}
\put(3.4,8.7){\makebox(0,0){$V_2$}}
\put(5.6,6.6){\makebox(0,0){$L_1$}}
\put(5.6,8.4){\makebox(0,0){$L_2$}}
\put(5.6,9.4){\makebox(0,0){$\tilde\Gamma$}}
\put(1.4,9.4){\makebox(0,0){$P_1^0$}}
\put(2.4,6.6){\makebox(0,0){$P_2^0$}}
\end{picture}}
\label{fig:evmt0}
}
\hspace{2cm}
\subfigure[Points on $\Gamma$ from $t=0$ to $4$]
{\unitlength=.3in{\def\arraystretch{1.0}
\begin{picture}(5,11)(0,2)
\put(1,3){\line(1,2){1}}
\put(1,3){\line(0,1){9}}
\put(1,12){\line(1,-2){1}}
\put(2,5){\line(0,1){5}}
\put(2,5){\line(1,1){1}}
\put(2,10){\line(1,-1){1}}
\put(3,6){\line(0,1){3}}
\put(1,3){\line(-1,-3){.3}}
\put(1,12){\line(-1,3){.3}}
\put(3,6){\line(1,0){3}}
\put(3,9){\line(1,0){3}}
\thicklines
\put(2,7){\circle{.2}}
\put(1,9){\circle*{.2}}
\put(2,6){\circle{.2}}
\put(1,7){\circle*{.2}}
\put(2,5){\circle{.2}}
\put(1,5){\circle*{.2}}
\put(3,6){{\circle{.2}}}
\put(1,4){{\circle*{.2}}}
\put(3,7){{\circle{.2}}}
\put(1,3){{\circle*{.2}}}
\put(5.6,9.4){\makebox(0,0){$\tilde\Gamma$}}
\put(1.5,3){\makebox(0,0){$P_1^4$}}
\put(3.5,7.){\makebox(0,0){$P_2^4$}}
\put(.5,4){\makebox(0,0){$P_1^3$}}
\put(3.3,5.6){\makebox(0,0){$P_2^3$}}
%
%
\put(.5,7.){\makebox(0,0){$P_1^1$}}
\put(2.4,6.){\makebox(0,0){$P_2^1$}}
\put(.5,5){\makebox(0,0){$P_1^2$}}
\put(2.4,4.7){\makebox(0,0){$P_2^2$}}
\put(.5,9.){\makebox(0,0){$P_1^0$}}
\put(2.4,7){\makebox(0,0){$P_2^0$}}
\end{picture}}
\label{fig:k1total}
}
\caption{Intersection of tropical plane curves $\Gamma$, $L_1^t$ and $L_2^t$ which realizes the eigenvector map $\phi$ (left; for $t=0$).
Each set $\{\bJ^t,\bW^t\}$ of values of the pBBS is mapped into the intersection points $P_1^t$ and $P_2^t$ of $\Gamma$, $L_1^t$ and $L_2^t$ by the tropical eigenvector map $\phi$ (right; for $t=0,1,\ldots,4$).
}
\end{figure}
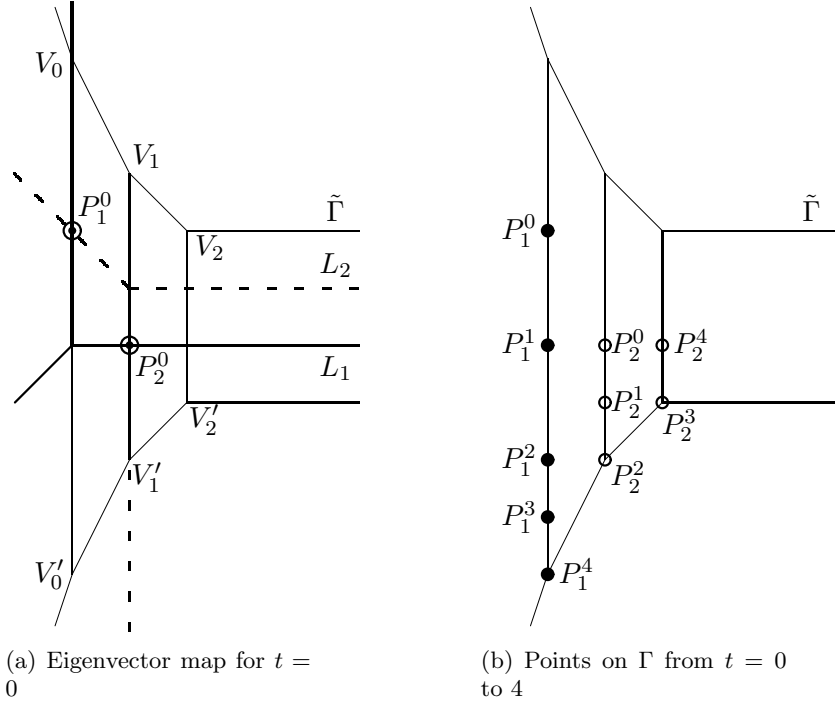

\begin{figure}[htbp]
\centering
\subfigure[Evolution from $t=0$ to $1$]
{\unitlength=.3in{\def\arraystretch{1.0}
\begin{picture}(5,11)(0,2)
\put(1,3){\line(1,2){1}}
\put(1,3){\line(0,1){9}}
\put(1,12){\line(1,-2){1}}
\put(2,5){\line(0,1){5}}
\put(2,5){\line(1,1){1}}
\put(2,10){\line(1,-1){1}}
\put(3,6){\line(0,1){3}}
\put(1,3){\line(-1,-3){.3}}
\put(1,12){\line(-1,3){.3}}
\put(3,6){\line(1,0){3}}
\put(3,9){\line(1,0){3}}
\thicklines
\put(1.0,5){{\line(0,1){8}}}
\put(1.0,5){{\line(-1,-2){.7}}}
\put(1.0,5){{\line(1,1){1}}}
\put(2.0,6){{\line(0,1){7}}}
\put(2.0,6){{\line(1,0){4}}}
\put(2,7){\circle*{.2}}
\put(1,9){\circle*{.2}}
\put(2,6){\circle{.3}}
\put(1,7){\circle{.3}}
\put(2,6){\circle*{.1}}
\put(1,7){\circle*{.1}}
\put(2,9){\circle{.2}}
\put(1,8){\circle{.2}}
\put(5.6,9.4){\makebox(0,0){$\tilde\Gamma$}}
\put(5.6,5.6){\makebox(0,0){$K_1^0$}}
\put(.5,9.){\makebox(0,0){$P_1^0$}}
\put(2.4,7){\makebox(0,0){$P_2^0$}}
\put(.5,7.){\makebox(0,0){$P_1^1$}}
\put(1.6,6.2){\makebox(0,0){$P_2^1$}}
\put(.5,8){\makebox(0,0){$Q_1^0$}}
\put(2.5,9){\makebox(0,0){$Q_2^0$}}
\put(.6,11.9){\makebox(0,0){$V_0$}}
\put(2.4,10.1){\makebox(0,0){$V_1$}}
\put(3.2,5.6){\makebox(0,0){$V_2^\prime$}}
\put(.6,5.2){\makebox(0,0){$U_0$}}
\put(2.4,6.3){\makebox(0,0){$U_1$}}
\end{picture}}
\label{fig:k1t0}
}
\hspace{2cm}
\subfigure[Evolution from $t=1$ to $2$]
{\unitlength=.3in{\def\arraystretch{1.0}
\begin{picture}(5,11)(0,2)
\put(1,3){\line(1,2){1}}
\put(1,3){\line(0,1){9}}
\put(1,12){\line(1,-2){1}}
\put(2,5){\line(0,1){5}}
\put(2,5){\line(1,1){1}}
\put(2,10){\line(1,-1){1}}
\put(3,6){\line(0,1){3}}
\put(1,3){\line(-1,-3){.3}}
\put(1,12){\line(-1,3){.3}}
\put(3,6){\line(1,0){3}}
\put(3,9){\line(1,0){3}}
\thicklines
\put(1.0,5){{\line(0,1){8}}}
\put(1.0,5){{\line(-1,-2){.7}}}
\put(1.0,5){{\line(1,1){1}}}
\put(2.0,6){{\line(0,1){7}}}
\put(2.0,6){{\line(1,0){4}}}
\put(2,6){\circle*{.2}}
\put(1,7){\circle*{.2}}
\put(2,5){\circle{.3}}
\put(1,5){\circle{.3}}
\put(2,5){\circle*{.1}}
\put(1,5){\circle*{.1}}
\put(2,10){\circle{.2}}
\put(1,10){\circle{.2}}
\put(5.6,9.4){\makebox(0,0){$\tilde\Gamma$}}
\put(5.6,5.6){\makebox(0,0){$K_1^1$}}
\put(.5,7.){\makebox(0,0){$P_1^1$}}
\put(1.6,6.2){\makebox(0,0){$P_2^1$}}
\put(.5,10){\makebox(0,0){$Q_1^1$}}
\put(2.5,10.2){\makebox(0,0){$Q_2^1$}}
\put(.5,5){\makebox(0,0){$P_1^2$}}
\put(2.4,4.7){\makebox(0,0){$P_2^2$}}
\end{picture}}
\label{fig:k1t1}
}
\\
\subfigure[Evolution from $t=2$ to $3$]
{\unitlength=.3in{\def\arraystretch{1.0}
\begin{picture}(5,11)(0,2)
\put(1,3){\line(1,2){1}}
\put(1,3){\line(0,1){9}}
\put(1,12){\line(1,-2){1}}
\put(2,5){\line(0,1){5}}
\put(2,5){\line(1,1){1}}
\put(2,10){\line(1,-1){1}}
\put(3,6){\line(0,1){3}}
\put(1,3){\line(-1,-3){.3}}
\put(1,12){\line(-1,3){.3}}
\put(3,6){\line(1,0){3}}
\put(3,9){\line(1,0){3}}
\thicklines
\put(1.0,4){{\line(0,1){9}}}
\put(1.0,4){{\line(-1,-2){.7}}}
\put(1.0,4){{\line(1,1){2}}}
\put(3.0,6){{\line(0,1){7}}}
\put(3.0,6){{\line(1,0){3}}}
\put(3,9){\circle{.2}}
\put(1,11){\circle{.2}}
\put(3,6){{\circle{.3}}}
\put(1,4){{\circle{.3}}}
\put(3,6){{\circle*{.1}}}
\put(1,4){{\circle*{.1}}}
\put(2,5){{\circle*{.2}}}
\put(1,5){{\circle*{.2}}}
\put(5.6,9.4){\makebox(0,0){$\tilde\Gamma$}}
\put(5.6,5.6){\makebox(0,0){$K_1^2$}}
\put(.5,5){\makebox(0,0){$P_1^2$}}
\put(2.4,4.7){\makebox(0,0){$P_2^2$}}
\put(.5,11){\makebox(0,0){$Q_1^2$}}
\put(3.4,9.4){\makebox(0,0){$Q_2^2$}}
\put(.5,4){\makebox(0,0){$P_1^3$}}
\put(3.4,6.4){\makebox(0,0){$P_2^3$}}
\end{picture}}
\label{fig:k1t2}
}
\hspace{2cm}
\subfigure[Evolution from $t=3$ to $4$]
{\unitlength=.3in{\def\arraystretch{1.0}
\begin{picture}(5,11)(0,2)
\put(1,3){\line(1,2){1}}
\put(1,3){\line(0,1){9}}
\put(1,12){\line(1,-2){1}}
\put(2,5){\line(0,1){5}}
\put(2,5){\line(1,1){1}}
\put(2,10){\line(1,-1){1}}
\put(3,6){\line(0,1){3}}
\put(1,3){\line(-1,-3){.3}}
\put(1,12){\line(-1,3){.3}}
\put(3,6){\line(1,0){3}}
\put(3,9){\line(1,0){3}}
\thicklines
\put(1.0,4){{\line(0,1){9}}}
\put(1.0,4){{\line(-1,-2){.7}}}
\put(1.0,4){{\line(1,1){2}}}
\put(3.0,6){{\line(0,1){7}}}
\put(3.0,6){{\line(1,0){3}}}
\put(3,8){\circle{.2}}
\put(1,12){\circle{.2}}
\put(3,6){{\circle*{.2}}}
\put(1,4){{\circle*{.2}}}
\put(3,7){{\circle{.3}}}
\put(1,3){{\circle{.3}}}
\put(3,7){{\circle*{.1}}}
\put(1,3){{\circle*{.1}}}
\put(5.6,9.4){\makebox(0,0){$\tilde\Gamma$}}
\put(5.6,5.6){\makebox(0,0){$K_1^3$}}
\put(1.5,3){\makebox(0,0){$P_1^4$}}
\put(3.5,7.){\makebox(0,0){$P_2^4$}}
\put(.5,4){\makebox(0,0){$P_1^3$}}
\put(3.3,5.6){\makebox(0,0){$P_2^3$}}
\put(.5,12){\makebox(0,0){$Q_1^3$}}
\put(3.5,8){\makebox(0,0){$Q_2^3$}}
\end{picture}}
\label{fig:k1t3}
}
\caption{Intersection of tropical plane curves $\Gamma$ and $K_1^t$ which is equivalent to the time evolution of the pBBS imposing the initial values \eqref{eq:ivpBBS}.
}
\end{figure}
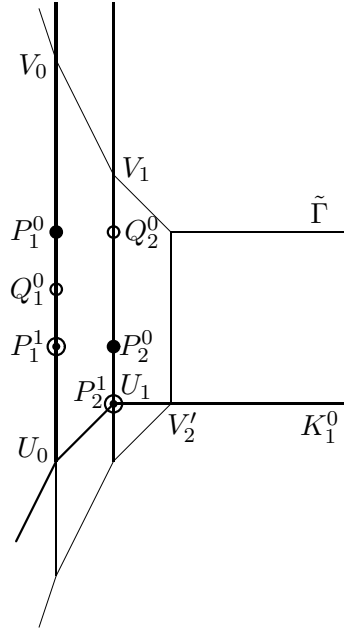
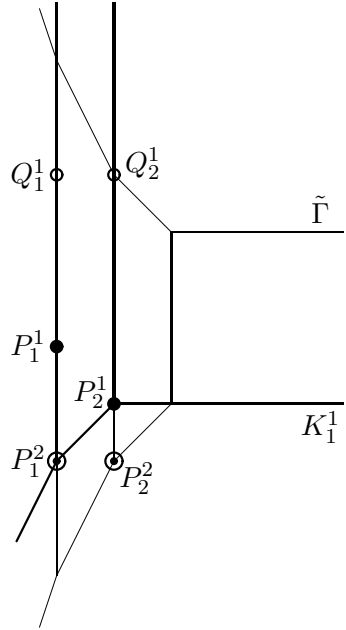
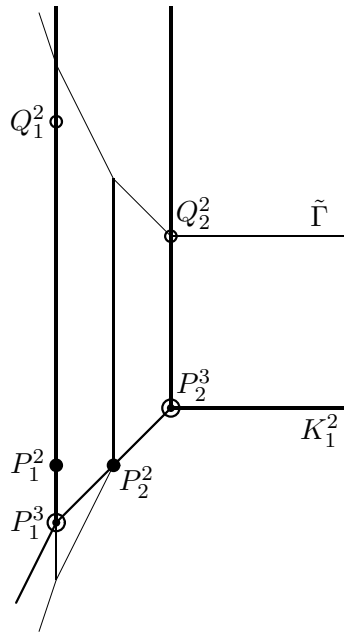
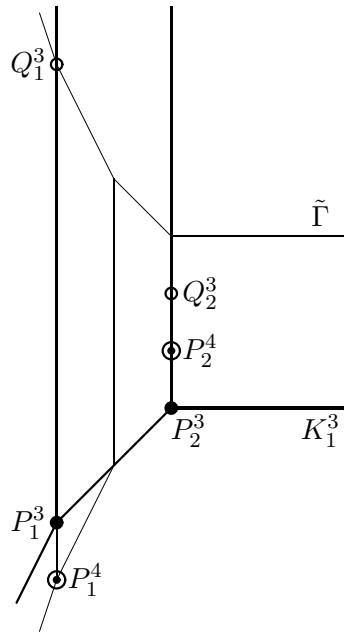

Now consider time evolution of the pBBS.
The tropical curve $K_1$ is given by the tropical polynomial
\begin{align*}
H_1(X,Y)
=
\left\lfloor
Y,
S_1C_2+2X,S_1C_1+X,S_1C_0
\right\rfloor,
\end{align*}
where
\begin{align*}
&S_1C_2
=
C_2(J_1-M,J_2,J_3;\bW)+M
=
J_1,
\\
&S_1C_1
=
C_1(J_1-2M,J_2,J_3;\bW)+2M
=
\left\lfloor
J_1+J_2,J_3+J_1,J_1+W_2
\right\rfloor,
\\
&S_1C_0
=
C_0(J_1-3M,J_2,J_3;\bW)+3M
=
J_1+J_2+J_3.
\end{align*}
For the initial values \eqref{eq:ivpBBS}, we obtain
\begin{align*}
&S_1C_2^0
=
J_1^0
=
3,
\\
&S_1C_1^0
=
\left\lfloor
J_1^0+J_2^0,J_3^0+J_1^0,J_1^0+W_2^0
\right\rfloor
=
\left\lfloor
5,4,5
\right\rfloor
=
4,
\\
&S_1C_0^0
=
J_1^0+J_2^0+J_3^0
=
6
\end{align*}
and hence
\begin{align*}
H_1^0(X,Y)
=
\left\lfloor
Y,
3+2X,4+X,6
\right\rfloor.
\end{align*}
Since we fix $g=2$, we do not need the second curve $K_2$ given by $H_2$ to realize the time evolution of the pBBS.

Figure \ref{fig:k1t0} shows the intersection of $K_1^0$ and $\Gamma$.
The conjugates ${Q_1^0}^\prime$ and ${Q_2^0}^\prime$ of the new intersection points $Q_1^0$ and $Q_2^0$ are $P_1^1$ and $P_2^1$, and these points correspond to $\bJ^1$ and $\bW^1$ through the eigenvector map $\phi$.
Figures \ref{fig:k1t1} -- \ref{fig:k1t3} show the intersection of $\Gamma$ and $K_1^t$ equivalent to the time evolution of the pBBS for $t=1,2,3$.

To obtain the intersection points concretely, we use the notion of stable intersection \cite{Vigeland04} and rational functions on $\Gamma$.
For example, in figure \ref{fig:k1t0}, we see that $\Gamma$ intersects $K_1^0$ at 5 vertices ($V_0, V_1, V_2^\prime\in\Gamma$ and $U_0, U_1\in K_1^0$) in the sense of stable intersection.
This fact suggests $V_0+V_1+U_0+U_1+V_2^\prime-V_0^\prime-2D^\ast\equiv0$ (mod $\mathcal{D}_l(\Gamma)$) (see \eqref{eq:addtroppic}).
We easily find that there exists a rational function on $\Gamma$ whose principal divisor is $P_1^0+P_2^0+Q_1^0+Q_2^0-V_0-V_1-U_0-U_1$.
This implies
\begin{align*}
P_1^0+P_2^0+Q_1^0+Q_2^0+V_2^\prime-V_0^\prime-2D^\ast\equiv0
\quad
(\mbox{mod $\mathcal{D}_l(\Gamma)$}).
\end{align*}
Thus the intersection points of $\Gamma$ and $K_1^0$ are $P_1^0$, $P_2^0$, $Q_1^0$, $Q_2^0$ and $V_2^\prime$.


\end{document}